\documentclass[aps,prl,showpacs,floatfix,twocolumn]{revtex4}

\usepackage[colorlinks=true,linkcolor=blue]{hyperref}

\usepackage{amsmath,bm}
\usepackage[dvipdfm]{graphicx}

\def\v2{\mbox{$v_2$}}

\begin{document}

%\draft
%
\hyphenation{author another created financial paper re-commend-ed Post-Script}

%\title{ Has the QCD Critical Point already been observed at RHIC ? }
\title{ Has the QCD Critical Point been Signaled by Observations at RHIC ? }
\author{ Roy~A.~Lacey} 
\author{ N.~N.~Ajitanand} 
\author{ J.~M.~Alexander}
\author{ P.~Chung}
\author{ W.G.~Holzmann} 
\author{ M.~Issah}
\author{A.~Taranenko}
\affiliation{ Department of Chemistry, State University of New York at Stony Brook, Stony 
Brook, NY 11794-3400, USA
} 
\author{P.~Danielewicz$^1$ and Horst St\"ocker$^2$}
\affiliation{$^1$National Superconducting Cyclotron Laboratory and
Department of Physics and Astronomy, Michigan State University, 
East Lansing, MI 48824-1321, USA \\
$^2$Institut f\"ur Theoretische Physik, Johann Wolfgang Goethe-Universit\"at 
             D–60438 Frankfurt am Main, Germany 
}
%
%\author{Horst St\"ocker}
%
%\affiliation{Institut f\"ur Theoretische Physik, J. W. Goethe-Universit\"at
%             Max von Laue Strasse 1, 60438 Frankfurt am Main, Germany}
%
\date{\today}

\begin{abstract}
The shear viscosity to entropy ratio ($\eta/s$) is estimated for the hot and 
dense QCD matter created in Au+Au collisions at RHIC ($\sqrt{s_{NN}}=200$~GeV). A very 
low value is found $\eta/s \sim 0.1$, which is close to the conjectured lower bound ($1/4\pi$). 
It is argued that such a low value is indicative of thermodynamic trajectories for the 
decaying matter which lie close to the QCD critical end point.
\end{abstract}

% \keywords{}
\maketitle

%\narrowtext 
%
%\section{Introduction}

	A primary objective for studying ultra-relativistic heavy ion collisions 
is to map out the accessible regions of the Quantum Chromo Dynamics (QCD) phase diagram.
The existence of a critical point -- an end point of the first
order chiral transition in this diagram -- has attracted considerable attention 
since it was first proposed \cite{Asakawa:1989bq,Barducci:1989wi}. 
Several theoretical techniques have been exploited to locate this critical 
end point (CEP) in the plane of temperature vs baryon chemical potential ($T, \mu_B$) 
i.e. the QCD phase diagram 
\cite{Fodor:2001pe,deForcrand:2003hx,Allton:2005gk,Gavai:2004sd,Philipsen:2005mj} . 
The properties of the CEP have also been sought via model calculations and 
the universality hypothesis \cite{Berges:1998rc,Halasz:1998qr,Stephanov:2004wx}. 
Nonetheless, experimental verification of the CEP constitutes a major 
current scientific challenge.

	The study of heavy ion collisions to search for the CEP was proposed 
several years ago \cite{Stephanov:1998dy}. Recently, a resurgence of experimental 
interest has centered around the prospects for successful experimental searches 
at Brookhaven's Relativistic Heavy Ion Collider (RHIC) \cite{CpWorkshop:2006,Stephans:2006tg} 
and CERN's Super Proton Synchrotron (SPS) \cite{Gazdzicki:1998vd,Gazdzicki:2005gs}. 
Indeed, a recent theoretical investigation involving two flavors of light dynamical 
quarks \cite{Gavai:2005mu} gives $T_c/T_{co} \sim 0.95$ and $\mu^c_B/T_{co} \sim 1.1$, 
where $T_{co} \sim 170$~MeV is the temperature for a cross-over to the quark gluon 
plasma (QGP) at zero baryon chemical potential and $T_c$ and $\mu^c_B$ are 
the temperature and the baryon chemical potential (respectively) at the critical end point.
It is widely believed that these values for $T_c$ and $\mu^c_B$ place the 
CEP in the range of observability for future energy scans at both RHIC and 
the SPS \cite{Stephans:2006tg}.  

\begin{figure}[tb]
 \includegraphics[width=1.0\linewidth]{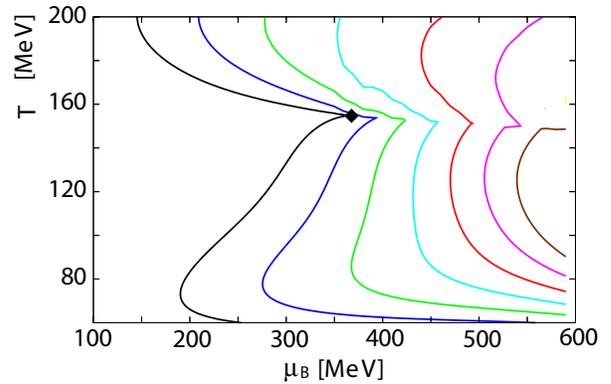}  % fig1.eps
 \caption{\label{fig1}
	(Color online) Results from a hydrodynamical calculation by Nonaka et al which 
assumes an EOS with a CEP \cite{Nonaka:2004pg}. Isentropic trajectories are 
shown for the case in which the CEP is located at ($T, \mu_B$) = (154.7 MeV, 367.8 MeV).
The value of $s/n_B$ for each of these trajectories (left to right) is 
100.0, 66.6, 50.0, 40, 33.3, 28.6 and 25. 
}
\end{figure}

	In recent hydrodynamical calculations \cite{Nonaka:2004pg,Kampfer:2005nt}, an 
interesting aspect of the CEP has been reported which could have important 
consequences for its detection. That is, it acts as an attractor for thermodynamic 
trajectories of evolving hot and dense QCD matter \cite{Stephanov:1998dy}.
This aspect of the CEP is illustrated in Fig.~\ref{fig1} where isentropic trajectories 
in the $(T, \mu_B)$ plane are shown for a broad range of values for the entropy 
per baryon $s/n_B$. Trajectories for the range $s/n_B \sim 100 - 50$, are clearly 
focused toward the CEP. Such a focusing is not observed for isentropic 
trajectories in similar model calculations which do not take explicit account
of the CEP in the equation of state (EOS). Instead, trajectories are merely 
shifted on the phase coexistence line \cite{Nonaka:2004pg}.
Given this aspect of the CEP, it is worthwhile to investigate whether or not 
the decay dynamics of the high energy density QCD matter created in heavy 
ion RHIC collisions show any signals for the CEP.

%Fig 1 :%%%

	Here, we argue that estimates for the ratio of viscosity to entropy 
density ($\eta/s$) made from RHIC measurements 
\cite{Teaney:2003kp,Gavin:2006xd,Lacey:2006qb}, indicate 
decay trajectories which are close to the CEP.
This implies that the prospects for constraining the location of the CEP in 
the $(T, \mu_B)$ plane are very good. Possibly, such constraints can be best achieved 
via a study of excitation functions with relatively large steps in collision energy
prior to focusing on a series of fine steps.

	There is strong experimental evidence for the creation of locally equilibrated 
nuclear matter at unprecedented energy densities, in heavy ion collisions at 
RHIC~\cite{Adcox:2004mh,Adams:2005dq,Back:2004je,Arsene:2004fa,
Gyulassy:2004zy,Muller:2004kk,Shuryak:2004cy,Heinz:2001xi}.
Jet suppression studies indicate that the constituents of this matter 
interact with unexpected strength and this matter is almost opaque to 
high energy partons \cite{Adler:2002tq,Adler:2005ee}. 
%
%Fig 2 :%%%

 \begin{figure}[tb]
 \includegraphics[width=1.0\linewidth]{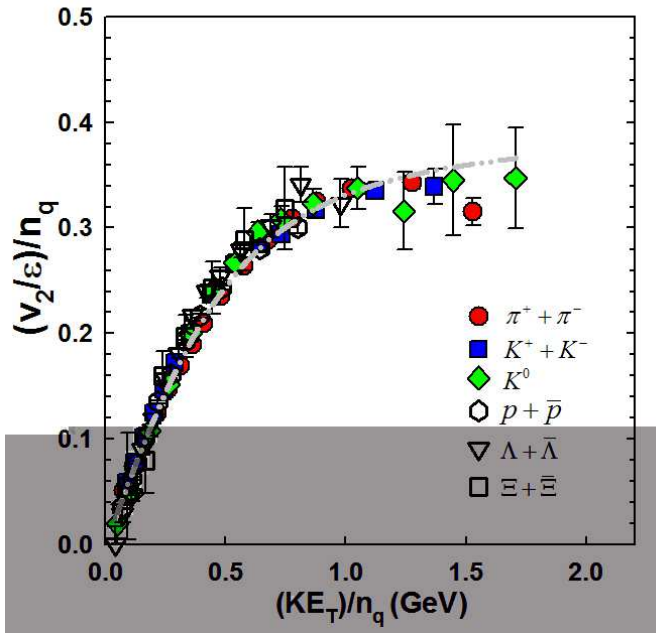}  % fig2.eps
 \caption{\label{fig2}
	(Color online) $v_2/\varepsilon n_q$ vs $KE_T/n_q$ for several identified particle species 
obtained in minimum bias Au+Au collisions. The data is taken from 
Refs.~\cite{Issah:2006qn,Adare:2006ti} and the references therein. The 
dashed-dotted line represents a fit to the data.
}
\end{figure}
Elliptic flow measurements \cite{Issah:2006qn,Adare:2006ti} validate the 
predictions of perfect fluid hydrodynamics for the scaling of the elliptic 
flow coefficient $v_2$ with eccentricity $\varepsilon$, system size and 
transverse kinetic energy KE$_T$ \cite{Csanad:2005gv,Csanad:2006sp,Bhalerao:2005mm}; 
they also indicate the predictions of valence quark number ($n_q$)
scaling \cite{Voloshin:2002wa,Fries:2003kq,Greco:2003mm},
suggesting that quark-like degrees of freedom are pertinent when elliptic 
flow develops \cite{Issah:2006qn,Adare:2006ti}. The result of such scaling
is illustrated in Fig.~\ref{fig2} where we plot $v_2/n_q\varepsilon$ vs KE$_T/n_q$; 
it shows that the relatively ``complicated" dependence of $v_2$ on 
centrality, transverse momentum, particle type and quark number can be 
scaled to a single function \cite{Issah:2006qn,Adare:2006ti}. 

	The expected suppression of flow due to 
the shear viscosity, as predicted by weak-coupling transport 
calculations is also not observed. In fact, only a rather small viscosity, 
$\eta/s \lesssim 0.1$, could be accommodated by the data in early 
estimates \cite{Teaney:2003kp}. 
This value of $\eta/s$, which is much lower than that obtained from weak 
coupling QCD \cite{Arnold:2000dr} or hadronic 
computations \cite{Muronga:2003tb}, has been interpreted as evidence 
that the quark gluon plasma (QGP) created in the early phase of RHIC 
collisions is more strongly coupled than expected \cite{Shuryak:2004cy,Hirano:2005wx}. 
%That is, a small value of $\eta/s$ implies strong coupling. 
An alternative interpretation is that the low value for $\eta/s$ could 
result from an anomalous viscosity $\eta_A$, arising from 
turbulent color magnetic and electric fields dynamically generated in 
the expanding quark-gluon plasma \cite{asakawa06}. 
That is, $\eta^{-1} = \eta^{-1}_A + \eta^{-1}_C$ and 
$\eta_A$ dominates over the collisional viscosity $\eta_C$.

\begin{figure}[tb]
 \includegraphics[width=1.0\linewidth]{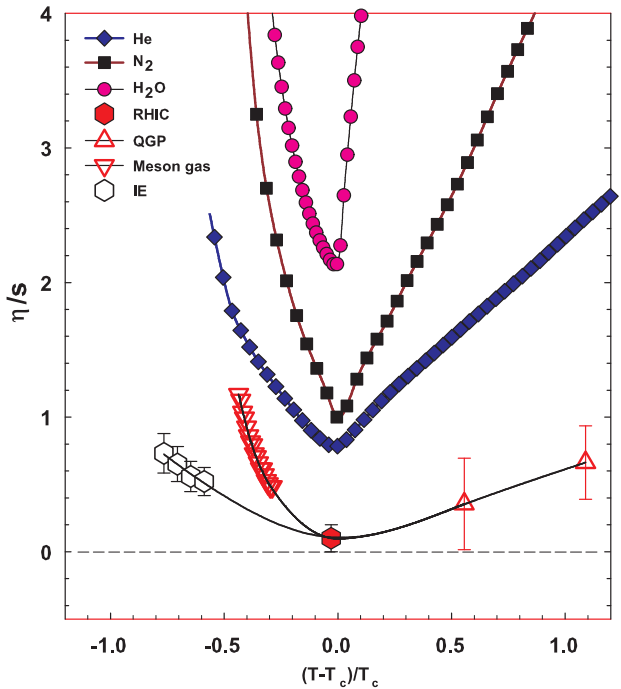}  % fig3.eps
 \caption{\label{fig3}
	(Color online) $\eta/s$ vs $(T-T_c)/T_c$ for several substances as indicated.
	The calculated values for the meson-gas have an associated error 
	of $\sim$ 50\%~\cite{Chen:2006ig}. 
	The lattice QCD value $T_c = 170$~MeV~\cite{Karsch:2000kv} 
	is assumed for nuclear matter. The lines are drawn to guide the eye.
}
 \end{figure}

	The ratio $\eta/s$ cannot be arbitrarily small because quantum mechanics limits
the size of cross sections via unitarity. A conjectured lower bound for $\eta/s$ is
$1/4\pi$, reached in the strong coupling limit of certain gauge 
theories \cite{Kovtun:2004de}. 
It has even been speculated that this lower bound holds for all 
substances \cite{Kovtun:2004de}. The temperature dependence of $\eta/s$ also provides 
invaluable insights. This is illustrated in Fig.~\ref{fig3} where 
we have plotted $\eta/s$ vs $(T-T_c)/T_c$ for molecular, atomic and nuclear 
matter. The data for He, N$_2$ and H$_2$O were obtained for their respective 
critical pressure. They are taken from Ref.~\cite{Csernai:2006zz} and the 
references therein. 
%data are taken from the National Institute of Standards and
%Technology (NIST) \cite{nist}. 
The calculated results shown for the meson-gas (for $T < T_c$) 
are obtained from chiral perturbation theory with free cross 
sections \cite{Chen:2006ig}. Those for the QGP (ie. for $T > T_c$)  
are from lattice QCD simulations \cite{Nakamura:2004sy}. 
The value $T_c \sim 170$ MeV is taken from lattice QCD 
calculations \cite{Karsch:2000kv}. 

	Figure~\ref{fig3} illustrates the observation that for 
atomic and molecular substances, the ratio $\eta/s$ exhibits a minimum of 
comparable depth for isobars passing in the vicinity of the liquid-gas
critical point \cite{Kovtun:2004de,Csernai:2006zz,Chen:2006ig}. 
When an isobar passes through the critical point (as shown in Fig.~\ref{fig3}), 
the minimum forms a cusp at $T_c$; when it passes below the critical point, 
the minimum is found at a temperature below $T_c$ (liquid side) but is 
accompanied by a discontinuous change across the phase transition. 
For an isobar passing above the critical point, a less pronounced 
minimum is found at a value slightly above $T_c$. 
The value $\eta/s$ is smallest in the vicinity of $T_c$ 
because this corresponds to the most difficult condition for the transport of 
momentum \cite{Csernai:2006zz}.

	Given the observations for atomic and molecular substances, one  
expects a broad range of trajectories in the $(T,n_B)$ or $(T,\mu_B)$ plane 
for nuclear matter, to show $\eta/s$ minima with a possible cusp at the 
critical point. The exact location of this point ( in the $(T,\mu_B)$ plane) 
is of course not known, and only coarse estimates of where it might lie 
are available. The open triangles in Fig.~\ref{fig3} show calculated values 
for $\eta/s$ along the $\mu_B=0$, $n_B=0$ trajectory. For $T < T_c$ the $\eta/s$ 
values for the meson-gas show an increase for decreasing values of $T$. 
For $T$ greater than $T_c$, the lattice results \cite{Nakamura:2004sy} 
indicate an increase of $\eta/s$ with $T$, albeit with large error bars.
These trends suggest a minimum for $\eta/s$. 

	The open hexagons in Fig.~\ref{fig3} show estimates for $\eta/s$ constrained 
by experimental data; they are for a trajectory which starts along $n_B=n_0$. 
The $\eta$ values are extracted from momentum transport data obtained 
in intermediate-energy (IE)
collisions \cite{pawel-1,Schmidt:1993ak,pawel-2}; values for $s$ were obtained 
from composite yields measured in the same energy and mass region.
The filled hexagon in  Fig.~\ref{fig3} shows a recent estimate for 
$\eta/s$ from high precision elliptic flow measurements in Au+Au 
collisions ($\sqrt{s_{NN}}=200$~GeV) at RHIC \cite{Lacey:2006qb}:
\[
	\frac{\eta}{s} \sim T \lambda_{f} c_s,
\]
where $T$ is the temperature, $\lambda_{f}$ is the mean free path and $c_s$ 
is the sound speed in the matter. 

	The temperature $T=160$~MeV was employed 
in a blast wave model generalized to include viscous corrections \cite{Teaney:2003kp}.
We estimate the value $T=165\pm 3$~MeV via a hydrodynamically inspired fit to the 
universally scaled $v_2$ data shown in Fig.~\ref{fig2}.
%
%Fig 3 :%%%
%
The functional form used for the fit is $I_1(w)/I_0(w)$, where $w=KE_T/2T$ 
and $I_1(w)$ and $I_0(w)$ are Bessel functions \cite{Csanad:2006sp}. 
For $c_s$, we use the recent observation that $v_2$ scales with eccentricity
and across system size to constrain the estimate 
$c_s = 0.35 \pm 0.05$~\cite{Issah:2006qn,Adare:2006ti,Bhalerao:2006tp}. 
The mean free path estimate $\lambda_f = 0.3 \pm 0.03 fm$ \cite{d_meson_flow}, is obtained
from an on-shell transport model simulation of the gluon evolution (in space 
and time) in Au+Au collisions at 200 GeV \cite{Xu:2004mz}. 
This parton cascade model includes pQCD $2 \leftrightarrow 2$ and 
$2 \leftrightarrow 3$ scatterings. These values for $\lambda_{f}$, $T$ and $c_s$ 
lead to the estimate $\eta/s \sim 0.09 \pm 0.015$ with an estimated systematic 
error of $ +0.1$ primarily due to the uncertainty in $\lambda_f$. 
This value for $\eta/s$ is in good agreement with the experimentally based 
estimates of Teaney and Gavin \cite{Teaney:2003kp,Gavin:2006xd} and the 
theoretical estimates of Gyulassy and Shuryak \cite{Hirano:2005wx,Gelman:2006xw}. 
However, it is in stark contrast to the predictions of pertubative 
QCD \cite{Teaney:2003kp,Arnold:2000dr}.

	The experimental and calculated $\eta/s$ values for nuclear matter 
shown in Fig.~\ref{fig3}, give strong indication for a minimum in the 
vicinity of $T_c$. This minimum, achieved in RHIC collisions, is rather close 
to the conjectured lower bound of $\eta/s = 1/4\pi$. We therefore conclude that 
it is compatible with the minimum expected if the hot and dense QCD matter 
produced in RHIC collisions follow decay trajectories which are close to the CEP. 
As discussed earlier, such trajectories could be readily followed if the CEP 
acts as an attractor for thermodynamic trajectories for the decaying 
matter \cite{Nonaka:2004pg,Kampfer:2005nt}. 
Here, it is noteworthy that the system has to spend a considerable 
portion of its dynamics in the region of low $\eta/s$ for that low value to 
have a significant impact on the reaction dynamics.  If the system gets 
momentarily, to a low $\eta/s$ and the rest of the dynamics takes place at 
high $\eta/s$, the momentary low value will have little impact.  An optimal 
situation would be if the system stays at low $\eta/s$ and then very quickly 
freezes out thereafter. That is, significant collective motion develops 
prior to the systems evolution toward higher $\eta/s$ values. 
 
	In summary we have used the comprehensive overall scaling of 
elliptic flow data to constrain a rudimentary experimental estimate for $\eta/s$ in Au+Au 
collisions at RHIC \cite{future}. A value close to the conjectured lower bound of $\eta/s = 1/4\pi$ 
is observed. We argue that such a low value for  $\eta/s$ is suggestive of decay 
trajectories which lie close to the QCD critical end point, 
because it acts as an attractor for constant entropy phase space trajectories of evolving 
hot and dense QCD matter. This suggests that the prospects for a successful search for the 
first order QCD phase transition and the possibility for establishing reliable
constraints (in the $(T,\mu_B)$ plane) for its associated critical end point 
are quite good. An initial measurement of excitation functions with 
coarse-grained collision energy steps may give sufficient information, or 
at least, give a better perspective on where to focus attention on fine steps.  

This work was supported by the US DOE under contract
DE-FG02-87ER40331.A008 and by the U.S. National Science 
Foundation under Grant No. PHY-0555893.
%
%\newpage
%
\bibliography{Critical_Point}
\end{document}